\documentclass[prmaterials,showpacs,reprint,twocolumn,amsmath,amssymb,aps,superscriptaddress]{revtex4-2}

\usepackage{graphicx}
\usepackage{dcolumn}
\usepackage{bm}
\usepackage{epstopdf}
\usepackage{color}
\usepackage{times}
\usepackage{url}
\usepackage{xspace}
\usepackage{xr}
\usepackage{chemformula}
\usepackage{float}

\newcommand{\pddmitFull}{Et$_{n}$Me$_{4-n}X$[Pd(dmit)$_2$]$_2$\xspace}
\newcommand{\pddmit}{$x$-[Pd(dmit)$_2$]$_2$\xspace}

\begin{document}
\title{$x$-[Pd(dmit)$_2$]$_2$ as a quasi-1D, scalene Heisenberg model}
\author{E.~P.~Kenny}
\email{elisekenny@gmail.com}
\affiliation{School of Mathematics and Physics, The University of Queensland, Brisbane, Queensland, Australia}
\author{A.~C.~Jacko}
\affiliation{School of Mathematics and Physics, The University of Queensland, Brisbane, Queensland, Australia}
\author{B.~J.~Powell}
\affiliation{School of Mathematics and Physics, The University of Queensland, Brisbane, Queensland, Australia}

\date{\today}

\begin{abstract}

From first principles, we calculate the Heisenberg interactions between neighboring dimers in several compounds within the \pddmitFull (Et = ethyl, Me = methyl, dmit = 2-thioxo-1,3-dithiole-4,5-dithiolate) family using an atomistic approach; with broken-symmetry density functional theory. In all materials, we find a scalene triangular model where the strongest exchange coupling along one crystallographic axis is up to three times larger than the others and that frustration further enhances this quasi-one-dimensionality. We calculate the N\'eel ordering temperature via the chain random phase approximation (CRPA). We show that the difference in the frustrated interchain couplings is equivalent to a single bipartite interchain coupling, favoring long-range magnetic order. We find that the N\'eel ordering temperatures are in good agreement with the experimentally measured values for most compounds.

\end{abstract}

\maketitle

\section{Introduction}

Charge transfer salts, especially the \pddmitFull (\pddmit) family, have been of intense interest for more than a decade. Geometric frustration and strong electron correlations lead to a wide range of exotic phenomena \cite{Powell2011, Kanoda2011}. All \pddmit compounds are Mott insulators at ambient pressure and low temperature, but changing the counter-ion ($X$ and $n$) leads to many different ground states.  Most salts exhibit anitferromagnetic order; for example, $X$-$n=$ As-0, As-1, As-2, N-0, and Sb-0 \cite{Nakamura2001, Kato2006, Kobayashi1998}. Others exhibit valence-bond order (P-1) \cite{Tamura2006,Shimizu2007}, charge order (Sb-2) \cite{Nakao2005, Tamura2005}, and spin liquid behavior (Sb-1) \cite{Itou2010, Itou2011, Kato2012, Zhou2017}. Those with antiferromagnetic order have been shown to exhibit unconventional superconducting behavior with the application of hydrostatic pressure or uniaxial strain \cite{Kato2004, Yamamoto2018}. Regardless of their ground state, most compounds in this family exhibit magnetic ordering at a temperature much lower than is expected based on the strength of their magnetic interactions. 

\begin{figure}
	\centering
	\includegraphics[width=0.7\columnwidth]{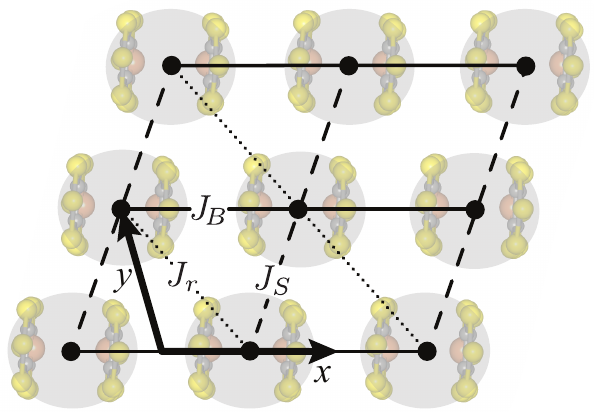}
	\caption{The intralayer geometry of \pddmit. Each layer forms a scalene (anisotropic) triangular lattice of dimers with interdimer exchange couplings $J_B$, $J_r$, and $J_S$ as shown. We find that $J_B$, along the dimer stacking direction, is the largest for all compounds in this paper.}
	\label{fig:geom}
\end{figure}

All compounds contain isomorphous layers Pd(dmit)$_2$ dimers separated by layers of counter-ions. The Pd(dmit)$_2$ dimers are arranged in a geometrically frustrated, scalene triangular lattice, differing only slightly between compounds. A single Pd(dmit)$_2$ layer is shown in  Fig. \ref{fig:geom}, the dimers form `stacks' along the (1,1,0)  lattice direction (along the horizontal in the figure). All of the materials studied here, except for P-1, are the so-called `solid crossing' bi-layer alternate where in successive layers the stacks are along the (1,1,0) and (1,$\bar{1}$,0) directions. In P-1 all layers are equivalent and stack along the (1,1,0)  lattice direction.

Their behavior above and below the Mott transition can be well described by a Hubbard model and, in the insulating phase, a Heisenberg model \cite{Powell2011}. In the insulating phase, there is one unpaired electron on each dimer. \textit{Ab initio} calculations on these compounds, including the parametrization of effective models, are difficult because of their chemical complexity. In these models, each spin site is an entire Pd(dmit)$_2$ monomer or dimer.

In this paper, we present a first-principles study of the magnetic interactions within several \pddmit compounds ($X$-$n=$ Sb-0, Sb-1, As-0, As-1, As-2, N-0, P-1). Using broken-symmetry density functional theory (BS-DFT), we find an effective Heisenberg model that is quasi-one-dimensional (one interaction, along the stacking direction, is much larger than the others). We then compare our model to experimental results using the chain random phase approximation (CRPA) together with the exact form of the dynamical susceptibility for a one-dimensional chain to calculate the N\'eel ordering temperature for each compound.

{In the past, these materials have been analysed and modeled through a quasi-two-dimensional picture \cite{Powell2011, Kanoda2011}. This paradigm stems from early calculations of dimer-dimer electron transfer integrals, $t_B$, $t_S$, and $t_r$ (c.f. Fig. \ref{fig:geom}) using an extended H\"uckel molecular orbital method, resulting in $t_B \approx t_S > t_r$ \cite{Kato2004}. This 2D picture was  reinforced by approximate  fits of the  2D triangular Heisenberg model to the magnetic susceptibility data \cite{Itou2008,Zheng2005}. However, more recent \textit{ab inito} investigations using more sophisticated methods like DFT have suggested that, for most compounds, $t_B > t_S \approx t_r$ \cite{Scriven2012, Nakamura2012, Jacko2013, Tsumuraya2013, Misawa2020, Kenny2020}. Our BS-DFT calculations  confirm and strengthen this result as we find that $J_B\gg J_s\approx J_r$. Hence, we argue that the \pddmit compounds should be understood as quasi-one-dimensional.}

Our calculations reveal that, in all materials, the strongest exchange coupling is along the dimer stacking direction ($J_B$; cf. Fig. \ref{fig:geom}). This leads to a quasi-one-dimensional Hamiltonian for all compounds studied here. Modeling these compounds via a quasi-1D approach allows us to calculate their N\'eel ordering temperatures analytically. Specifically, we use the chain random phase approximation (CRPA) around the large $J_B$ limit, starting from the exact form for the one-dimensional magnetic susceptibility of a Heisenberg spin-1/2 chain and treating interchain interactions via the RPA \cite{Schulz1996, Bocquet2001}. In the case of an isosceles triangular lattice, the interchain interactions are perfectly frustrated. This suppresses ordering at any temperature \cite{Bocquet2001, Kenny2019} within the CRPA. In \pddmit, we find that the anisotropy in the interchain coupling leads to an effective unfrustrated interchain interaction, given by the difference of the interchain couplings ($\delta J_y=J_r\!-\!J_S$). 

{Previous studies of \pddmit have parameterized tight-binding models on the basis of band structure calculations by fitting to models or via Wannier functions centered on monomers or dimers \cite{Itou2008, Powell2011, Nakamura2012, Scriven2012, Tsumuraya2013, Jacko2013, Misawa2020}. Extracting tight-binding parameters from band structure calculations relies on Kohn-Sham eigenvalues. These do not correspond to nature, but are rather an internal DFT device for calculating the total density \cite{KohnSham1965}. Kohn-Sham eigenvalues often poorly reproduce energy differences, even in weakly correlated materials \cite{Jones1989, Perdew1985}, and dramatically fail in strongly correlated materials \cite{Adler2018}. For example, in Sb-1, the Kohn-Sham band structure is metallic \cite{Scriven2012, Tsumuraya2013, Jacko2013, Nakamura2012, Powell2011} rather than insulating, as in experiment \cite{Powell2011,Kanoda2011}. In contrast, BS-DFT makes use of ground-state energy differences, which have a formal basis in DFT (the Hohenberg-Kohn theorem \cite{Hohenberg1964}) and are much more accurate in DFT than the Kohn-Sham eigenvalues. Moreover, it allows us to directly determine the Heisenberg exchange interaction, rather than extracting the magnetic behavior from tight-binding \cite{Scriven2012, Jacko2013, Tsumuraya2013} or Hubbard models \cite{Nakamura2012, Misawa2020}, where the electron-electron interaction parameters are often poorly known.}

The Heisenberg exchange interaction, $J_{ij}$, mainly arises from two different physical contributions -- superexchange (SE) and direct exchange (DE), $J_{ij} = J^\mathrm{SE}_{ij} +J^\mathrm{DE}_{ij}$. In molecular crystals such as \pddmit, $J^\mathrm{SE}_{ij}$ is often the largest term \cite{Nakamura2012, Kenny2020, Misawa2020}. It arises from virtual hopping processes and usually favors antiferromagnetism \cite{Anderson1950}. This interaction is encompassed by the Hubbard model. $J^\mathrm{DE}_{ij}$ arises from the antisymmetry of electron wavefunctions. It favors ferromagnetism and depends mostly on the distance between sites. Superexchange and direct exchange are usually of opposite sign; a significant direct exchange can counteract the superexchange interaction and lower the magnitude of $J_{ij}$. Estimates of $J_{ij}$ from H\"uckel and DFT band structure calculations typically neglect $J^\mathrm{DE}_{ij}$. However, studies that have calculated $J^\mathrm{DE}_{ij}$ directly have found that its magnitude is significant in these compounds \cite{Nakamura2001, Kenny2020, Misawa2020}. For this reason, studies that take direct exchange into account are bound to be more successful at accurately modeling experimental behaviors in \pddmit. Since we directly calculate the total $J_{ij}$ in this work, all its contributions (including superexchange and direct exchange) are present in our modeling.

\section{Parametrization of Heisenberg Model with BS-DFT}

We directly parameterize a Heisenberg model,
	\begin{equation}
	\mathcal{H}=\sum_{ ij} J_{ij}\ \bm{S}_i\cdot\bm{S}_j,
	\label{eq:HeisenHam}
	\end{equation} 
where $\bm{S}_i$ is the spin operator on the $i$th dimer and $J_{ij}$ are the exchange coupling constants. 
	
The exchange couplings, $J_{ij}$, are calculated as the energy difference between specific spin states of each tetramer in the compound -- using BS-DFT along with the Yamaguchi spin decontamination procedure. In this approach \cite{Noodleman1981, Mouesca, Yamaguchi1988},
	\begin{equation}
	J_{ij}=2\frac{E^\mathrm{BS}_{ij}-E^\mathrm{T}_{ij}}{\langle S^2\rangle^\mathrm{BS}_{ij}-\langle S^2\rangle^\mathrm{T}_{ij}},
	\label{eq:BS}
	\end{equation}
where $E^\mathrm{T}_{ij}$ is the triplet energy of the isolated tetramer (two neighboring dimers, $i$ and $j$) and $E^\mathrm{BS}_{ij}$ is the energy of the broken-symmetry state, where the unpaired spins on each dimer are misaligned. $\langle S^2\rangle^\mathrm{BS}_{ij}$ and $\langle S^2\rangle^\mathrm{T}_{ij}$ are the corresponding expectation values of the spin operator, $S^2$. The coordinates for each tetramer included two Pd(dmit)$_2$ dimers and the six closest counter-ions. Calculations were performed in Gaussian09 \cite{g09} with the uB3LYP functional \cite{B3LYPa,B3LYPb} and using the LANL2DZ \cite{Dunning1977, Hay1985a, Wadt1985, Hay1985b}  (for Pd, Sb, As, and Cs) and 6-31+G* \cite{6+31G_star} basis sets. We included the six nearest cations to each Pd(dmit)$_2$ tetramer; benchmarking revealed that the calculated exchange interactions are well converged at this cluster size. We used a collection of experimental crystal structures \cite{Yugo}.

\begin{table}
	\centering
	\begin{tabular}{c|cccc}
		Compound ($X$-$n$) & $J_B$ (K) & $J_r$ (K) & $J_S$ (K) & $J_{z}$ (K) \\
		\hline 
		\hline
		Sb-0     & 320 & 145  & 86  & -0.01  \\
		Sb-1     & 382 & 129  & 111  & 0.06 \\
		As-0     & 370 & 110  & 116 & -0.02 \\
		As-1     & 353 & 129  & 100 & $<0.01$\\
		As-2     & 374 & 130  & 98  & -0.01\\
		N-0     & 352  & 131  & 143  & 0.04  \\
		P-1      & 499 & 195  & 148  & 5       
	\end{tabular}
	\caption{Our BS-DFT Heisenberg exchange interactions of \pddmitFull  (the compounds are listed here according to their values of $X$ and $n$). $J_B$ is much larger than $J_{r}$ and $J_{S}$ in all materials. The interlayer exchange coupling, $J_z$, is very small. See Figure \ref{fig:geom} for the directions of the other couplings in the lattice. }
	\label{tab:Js}
\end{table}

\begin{figure}
	\centering
	\includegraphics[width=\columnwidth]{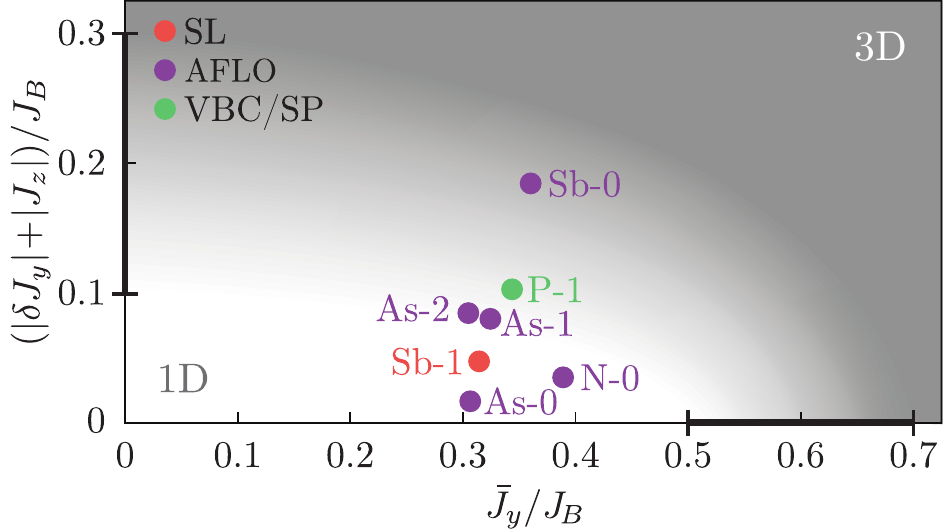}
	\caption{BS-DFT results for the frustrated inter-stack interaction, $\bar{J_y}=\frac{1}{2}\left(J_S\!+\!J_r\right)$, and the unfrustrated $|\delta J_y|\!+\!|J_z|=|J_r\!-\!J_S|\!+\!|J_z|$, in units of the coupling along the stack direction, $J_B$ (see Figure \ref{fig:geom}). The marks on the axes indicate cut-offs for the validity of the CRPA for each interaction. The upper bounds indicate the most optimistic value in the literature, whereas the lower bounds indicate the most pessimistic value \cite{Weng2006, Yunoki2006, Hayashi2007, Pardini2008, Jiang2009, Heidarian2009, Tay2010, Yasuda2005}. The shading is a guide to the eye based on these bounds. The colours differ based on the ground state of the compound according to the experimental literature; antiferromagnetic long-range order (AFLO), valence-bond crystal (VBC) or spin-Peierls (SP), and spin liquid (SL). }
	\label{fig:BS-DFTplot}
\end{figure}

Our BS-DFT calculations reveal three significant, antiferromagnetic, nearest-neighbor couplings shown in Table \ref{tab:Js} and Fig. \ref{fig:geom}. The largest exchange coupling $J_B$ (along the stacking direction in Fig. \ref{fig:geom}) is significantly larger than the others in all cases. Hence, our Heisenberg model is quasi-one-dimensional. For the two smaller (interchain) couplings, we make a change of variables to the average of the interchain couplings, $\bar{J_y}=\frac{1}{2}\left(J_S\!+\!J_r\right)$, and their difference, $\delta J_y=J_r\!-\!J_S$. Figure \ref{fig:BS-DFTplot} shows a plot of $\bar{J_y}$ and $|\delta J_y| + |J_z|$ for each compound.

The interlayer couplings ($J_z$ -- perpendicular to the axes in Figure \ref{fig:geom}) are all less than $0.01J_B$. However, they are unfrustrated and therefore have a non-negligible contribution to the magnetic behavior. P-1 has the largest $J_z$ by far, 5\,K, which is interesting, since it is the only compound purported to have a valence-bond crystal (VBC) or spin-Peierls (SP) ground state \cite{Tamura2006,Shimizu2007, Powell2011}. The quasi-1D picture naturally gives rise to a spin-Peierls ground state \cite{Cross1979}, whereas a 2D picture would suggest a VBC. Experimentally these are difficult to distinguish. The important difference is that, in a SP distortion,  spin-phonon coupling and lattice distortion are  essential ingredients of the mechanism; whereas the VBC is a fundamentally electronic effect and any lattice distortion is parasitic and driven by the spin-phonon coupling only on the lowering of the symmetry of the electrons. It is therefore interesting to note that the VBC/SP ground state is only observed in P-1 -- the only \pddmit material that does \textit{not} form the bilayer solid crossing structure. A lattice distortion would be strongly energetically disfavored in the solid crossing structure as there is an inherent elastic frustration between distortions along the (1,1,0) and (1,$\bar{1}$,0) directions in alternate layers \cite{Powell2011}. Hence, this elastic distortion may suppress the VBC/SP phase. This suggests that there is an SP distortion rather than a VBC state in P-1 and thus that the \pddmit materials are quasi-1D.

In a triangular, quasi-1D lattice there are two interesting limits to consider: when the lattice becomes perfectly frustrated, $\delta J_y$ and $J_z\rightarrow 0$, and when there is no geometrical frustration, $J_S\rightarrow 0$ or $J_r\rightarrow 0$. In the first case, the model becomes perfectly isosceles with two equal interchain couplings, $J_S=J_r=\bar{J_y}$. {Quantum Monte Carlo, exact diagonalization, density matrix renormalization group (DMRG), and other numerical calculations have shown that this model exhibits quasi-one-dimensional behavior for $\bar{J_y}/J_B<0.7$  \cite{Weng2006, Yunoki2006, Hayashi2007, Pardini2008, Jiang2009, Heidarian2009, Tay2010}. In the second case, we have a cuboidal lattice with interchain couplings $|\delta J_y|$ and $J_z$. This model, studied extensively by Schulz \cite{Schulz1996}, exhibits quasi-one-dimensional behavior for $(|\delta J_y|\!+\!|J_z|)/J_B<0.3$ according to RPA calculations \cite{Yasuda2005}. }These limits are indicated on the axes of Figure \ref{fig:BS-DFTplot}, where we find that the quasi-1D nature of the exchange couplings is evident in all compounds. Our calculations reveal that $\bar{J_y}$ is always below $0.4J_x$ and the sum of the unfrustated interactions, $|\delta J_y|+|J_z|$, is below $0.1J_x$ -- only Sb-0 is anywhere near these limits, as it is less strongly frustrated than the other materials. We conclude that all compounds can be described by quasi-one-dimensional models.

The results of our DFT calculations agree well with previous work. Accounting for the fact that $J_{ij} \sim t_{ij}^2$, where $t_{ij}$ is the tight-binding transfer integral, the couplings calculated in other systematic studies of \pddmit also reflect a strongly quasi-1D character. If we use the superexchange in the large $U$ limit, 
	$\bar{J}_y^\mathrm{SE}\approx \frac{4}{U}\frac{t_S^2+t_r^2}{2}$,
	$\delta J_y \approx \frac{4}{U}(t_S^2+t_r^2)$
	and $J_B^\mathrm{SE}\approx 4t_B^2/U$, 
	which leads to $J_B^\mathrm{SE}/\bar{J}_y^\mathrm{SE}\sim 2t_B^2/(t_r^2 + t_S^2)$, the values found by \citeauthor{Tsumuraya2013} \cite{Tsumuraya2013} and \citeauthor{Misawa2020} \cite{Misawa2020} all fit well within the 1D region in Fig. \ref{fig:BS-DFTplot}.

\section{Calculation of N\'eel temperature with the CRPA}

The CRPA expression for the three-dimensional dynamical magnetic susceptibility of a lattice of weakly coupled chains is \cite{Scalapino1975, Schulz1996, Essler1997, Bocquet2001}
\begin{equation}
\chi_\mathrm{CRPA}\left(\omega,\bm{k}, T\right)=\frac{\chi_\mathrm{1D}(\omega,k_x, T)}{1-2\tilde{J}_\perp\left(\bm k\right)\chi_\mathrm{1D}(\omega,k_x, T)},
\label{eq:3d_chi_pddmit}
\end{equation}
where $\chi_\mathrm{1D}(\omega,k_x, T)$ is the dynamical susceptibility for a single Heisenberg chain, $\tilde{J}_{\perp}(\bm k)$ is the Fourier transform of the interchain coupling, and $\bm{k}\!=\!\left(k_x,k_y,k_z\right)$ is the crystal momentum along the axes in Figure \ref{fig:geom} in units of the inverse lattice spacing.

\begin{widetext}
We find that
\begin{equation}\label{eq:Jperp}
\begin{split}
\tilde{J}_{\perp}(\bm k) &= J_r\cos\left(k_{y}-\frac{k_x}{2}\right) + J_S\cos\left(k_{y}+\frac{k_x}{2}\right) + J_z\cos\left(k_z\right)\\
&=\left(\bar{J_y} + \delta J_y\right)\cos\left(k_{y}-\frac{k_x}{2}\right) 
+ \left(\bar{J_y} -\delta J_y\right)\cos\left(k_{y} + \frac{k_x}{2}\right) + J_z\cos\left(k_z\right),
\end{split}
\end{equation}
where, in the second line, we have used our change of variables defined above.

The dynamical susceptibility for a single Heisenberg chain around $k_x=k_0+\pi\approx\pi$ has been calculated from a combination of the Bethe ansatz and field theory techniques \cite{Bethe1931,  Schulz1983, Schulz1986,  Barzykin2000, TsvelikBook};
	\begin{equation}
	\chi_\mathrm{1D}(\omega,k_0, T) 
	=-\frac{\sqrt{\ln({\Lambda}/{T})}}{2t(2\pi)^{3/2}}
	\frac{\Gamma\left(\frac{1}{4}-i\frac{\omega-u k_0}{4\pi T}\right)}{\Gamma\left(\frac{3}{4}-i\frac{\omega-u k_0}{4\pi T}\right)}\frac{\Gamma\left(\frac{1}{4}-i\frac{\omega+u k_0}{4\pi T}\right)}{\Gamma\left(\frac{3}{4}-i\frac{\omega+u k_0}{4\pi T}\right)},
	\label{eq:chain_chi}
	\end{equation}
where $k_0=k_x-\pi$, $\Gamma(x)$ is the Euler gamma function, $u=\frac{\pi}{2}J_{x}b_0$ is the spin velocity, $b_0$ is the interdimer separation along the direction in Fig. \ref{fig:geom}, 
and $\Lambda\simeq24.27$  \cite{Barzykin2001}. 
\end{widetext}

We determine the N\'eel ordering temperature, $T_N$, by considering the condition for a zero frequency pole in Eq. \ref{eq:3d_chi_pddmit}. This occurs when
\begin{equation}
2\tilde{J}_{\perp}\left(\bm{k}\right)\chi_\mathrm{1D}(0,k_x,t)|_{T=T_N}=1.\label{eq:instab}
\end{equation} 
This instability will occur at the maximum of $\tilde{J}_{\perp}\left(\bm{k}\right)\chi_\mathrm{1D}(0,k_x,T)$. The presence of interchain couplings will shift this maximum to an incommensurate wavenumber, with the resulting order occurring at $k_x=\pi+k_0$. In this case, we find the maximum occurs when $k_{y}=\mathrm{arctan2}\left(-\delta J_y\cos(k_0/2),(2\bar{J_y}\sin(k_0/2))\right)$ where $\mathrm{arctan2}(y,x)$ is the two-argument arctangent, which returns the angle for the point ($x,y$) defined positively from the $x$-axis. A numerical solution of this condition using our BS-DFT results for $\bar{J_y}$, $\delta J_y$, and $J_z$, gives the N\'eel temperatures shown in Table \ref{tab:temps}. 

\begin{table}[h!]
	\begin{center}
		\begin{tabular}{c|cccc}
			Compound ($X$-$n$) & Calc. $T_N$\,(K) & Expt. $T_N$\,(K)& State\\ 
			\hline 
			\hline
			As-0 & 8.7 & 35        & AFLO \cite{Nakamura2001}\\ 
			As-1 & 37  & 23         & AFLO \cite{Kato2006}\\
			As-2 & 41  &  18         & AFLO \cite{Nakamura2001}\\ 
			N-0  & 17  & 12          & AFLO \cite{Kobayashi1998} \\
			Sb-0 & 70 & 18          & AFLO \cite{Nakamura2001}\\
			Sb-1 & 24 & $<$\,0.02 & SL  \cite{Itou2010}\\
			P-1   & 118 & 21          & VBC/SP \cite{Tamura2006,Shimizu2007}
		\end{tabular}
		\caption{ Calculated N\'eel temperatures of \pddmitFull based on nearest-neighbor coupling parameters in Table \ref{tab:Js} compared with experimental measurements. `State' refers to the ground state of the compound according to the experimental literature; antiferromagnetic long-range order (AFLO), valence-bond crystal (VBC) or spin-Peierls (SP), and spin liquid (SL).  }\label{tab:temps}
	\end{center} 
\end{table}

For all compounds in Table \ref{tab:temps}, we found $k_0<10^{-12}$. Taking the limit $k_0\rightarrow0$ in Eq. \ref{eq:instab} returns
\begin{equation}
T_N=0.5558\left(|J_z| + |\delta J_y|\right)\sqrt{\log\left(\frac{\Lambda}{T_N}\right)}.\label{eq:cube_form}
\end{equation} 
This takes the same form as the prediction for coupled chains with unfrustrated, bipartite interchain couplings of magnitude $\delta J_y$ along the $y$-axis and $J_z$ along the $z$-axis \cite{Schulz1996} -- demonstrating that $\delta J_y$ simply acts as an unfrustrated coupling, while the frustrated contribution $\bar{J_y}$ has no effect (since Eq. \ref{eq:cube_form} is independent of $\bar{J_y}$).

\begin{figure}
	\begin{center}
		\includegraphics[width=\columnwidth]{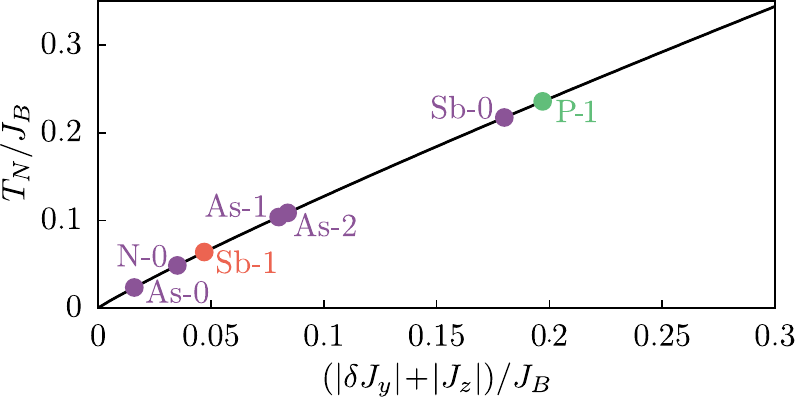}
		\caption{ Relationship between $T_N$ and the sum of the unfrustrated couplings, $|\delta J _y|+|J_z|$. Numerical calculations from Table \ref{tab:temps} are shown as labeled points with the same coloring as in Figure \ref{fig:BS-DFTplot}. The black line is Eq. \ref{eq:cube_form}, which passes perfectly through the numerical results. }
		\label{fig:tnVdelta}
	\end{center}
\end{figure}

Figure \ref{fig:tnVdelta} shows a plot of the N\'eel temperature from Eq. \ref{eq:cube_form} with the numerical results for each material indicated as points. The numerical results match those from Eq. \ref{eq:cube_form} perfectly. Comparing our results with experiment in Table \ref{tab:temps}, our estimates of $T_N$ are, mostly, not far from their experimental counter-parts. In general, our $T_N$ values are consistently higher than the experimental value. For example, if our model was qualitatively correct, the spin liquid candidate Sb-1, would have $T_N \approx 0$ and we do find a very small (but non-zero) $T_N$ for Sb-1. However, we find even lower N\'eel temperatures for As-0 and N-0. Our overestimation of $T_N$ could point to the importance of higher-order spin processes, such a ring-exchange \cite{Holt2014, Merino2014, Motrunich2005, Kenny2020} and that this has different magnitudes in different materials -- consistent with the different pressures/strains required to drive different materials metallic/superconducting. Nevertheless, a theory based on first principles calculations that gives quantitatively reasonable predictions for organic charge transfer salts is a significant advance.

\section*{Conclusions}

We have used BS-DFT, an atomistic approach, to parameterize a Heisenberg model for several materials in the \pddmitFull family. This revealed a frustrated scalene triangular lattice where the largest coupling along the stacking direction is nearly three times larger than the others. We showed that, in the relevant quasi-one-dimensional limit, the difference in the interchain coupling acts identically to an unfrustrated interchain coupling and favors long-range magnetic order. This is the role of geometric frustration in a quasi-1D triangular lattice; the effective interchain coupling, which is the main driver for magnetic order, is reduced significantly due to competing interactions. We calculate the N\'eel temperatures in this picture and find that they are similar to experimental values, but are over-estimates in most cases. This could indicate the importance of higher-order spin processes, such a ring-exchange, in the \pddmit family. 

Otherwise, treating these compounds as quasi-1D is consistent with the existing experimental literature and provides a natural explanation of why only P-1 has a SP distortion. In this picture, geometrical frustration and strong electron correlations still play a large role. In particular frustration increases the effective one-dimensionality ($\delta J_y \ll J_r$, $J_s$).
	
More broadly, the demonstration that we can get reason values for the N\'eel temperatures in the \pddmit materials  brings us close to achieving the long-held goal of making quantitative predictions for electronic phenomena in strongly correlated electron materials.

\section*{acknowledgements}

We thank Amie Khosla and Ross McKenzie for helpful conversations. 
This work was supported by the Australian Research Council through Grants No. DP160100060 and DP181006201.

\bibliographystyle{apsrev4-1}

\end{document}